\documentclass[review,onefignum,onetabnum]{siamart171218}

\usepackage{lipsum}
\usepackage{amsfonts}
\usepackage{graphicx}
\usepackage{epstopdf}
\usepackage{algorithmic}
\usepackage{booktabs}
\ifpdf
  \DeclareGraphicsExtensions{.eps,.pdf,.png,.jpg}
\else
  \DeclareGraphicsExtensions{.eps}
\fi

\newsiamremark{remark}{Remark}
\newsiamremark{hypothesis}{Hypothesis}
\crefname{hypothesis}{Hypothesis}{Hypotheses}
\newsiamthm{claim}{Claim}

\headers{Exploring usability of Reddit in data science}{J. Sawicki, M. Ganzha, M. Paprzycki, A.B\u{a}dic\u{a}}

\title{Exploring usability of Reddit in data science and knowledge processing
}

\author{Jan Sawicki\thanks{Warsaw University of Technology, 
  Department of Mathematics and Information Sciences,
  \email{jan.sawicki2.dokt@pw.edu.pl}, 
  .}
\and Maria Ganzha\thanks{Warsaw University of Technology, 
  Department of Mathematics and Information Sciences
  \email{m.ganzha@mini.pw.edu.pl}.}
\and Marcin Paprzycki\thanks{Systems Research Institute Polish Academy of Sciences
  \email{marcin.paprzycki@ibspan.waw.pl}.}
\and Amelia B\u{a}dic\u{a}\thanks{University of Craiova, Department of Statistics and Business Informatics \email{amelia.badica@edu.ucv.ro}.}
}

\usepackage{amsopn}

\makeatletter
\newcommand*{\addFileDependency}[1]{
  \typeout{(#1)}
  \@addtofilelist{#1}
  \IfFileExists{#1}{}{\typeout{No file #1.}}
}
\makeatother

\ifpdf
\hypersetup{
  pdftitle={Exploring usability of Reddit in data science},
  pdfauthor={J. Sawicki, M. Ganzha, M. Paprzycki, A.B\u{a}dic\u{a}}
}
\fi

\begin{document}

\maketitle

\begin{abstract}

This contribution argues that Reddit, as a massive, categorized, open-access dataset, is a useful data source, for ``almost any topic''. Hence, it can be used in data science, e.g. for knowledge exploration. This statement is backed-up with presented analysis, based on 180 manually annotated papers, related to Reddit itself, and data acquired from popular databases of scientific papers. Finally, an open source tool is introduced, which provides an easy access to Reddit resources, and an exploratory data analysis of how Reddit covers selected topics. These functions can be used as a prelude analysis to a broader exploration of Reddit's applicability.
\end{abstract}



\section{Introduction}\label{sec:introduction}

Recently, social networks and content sharing networks became popular repositories of data, used for information and knowledge processing (especially for information retrieval). The aim of this work is to explore the usability of Reddit as a data source. In this context, we present a review of scientific literature about Reddit itself, its presence in scientific databases, and elaborate its ``topical coverage''. Moreover, for the latter study, a specialized tool (\textit{Reddit-TUDFE})' is introduced, which allows for fast check of Reddit coverage of a selected topic. The key contributions of this work are answers to the following research question (RQs):
\begin{itemize}

\item 
\textbf{RQ1}: What are the most popular methods to acquire Reddit data? (do they allow capturing graph networks\footnote{Here, graph networks are of special interests, because it can be observed that large number of methods of data extraction and analysis are focused on application of graph theory.})

\item
\textbf{RQ2}: What problems are the most researched when using Reddit as a dataset?

\item
\textbf{RQ3}: How does Reddit usage in data science change over time? Is it declining or is it increasing?

\item
\textbf{RQ4}: Are there any popular topics that are not (substantially) covered on Reddit? 

\item
\textbf{RQ5}: Is Reddit used as a single dataset, or with datasets from other online platforms?\footnote{Answer to this question is crucial to establish (suggest) additional datasets, which could/should be used with Reddit.}

\end{itemize}

These questions are essential for further planned research and positive answer would mean that Reddit is a proper choice for proceeding with the project of information retrieval about popular trends, using graph databases and complex networks. Moreover, positive answers would indicate that Reddit may be a competitor (or a companion) to explorations based on more popular data sources, like Twitter.

\section{What is Reddit}

\begin{figure}[htpb]
    \centering
    \includegraphics[width=0.5\linewidth]{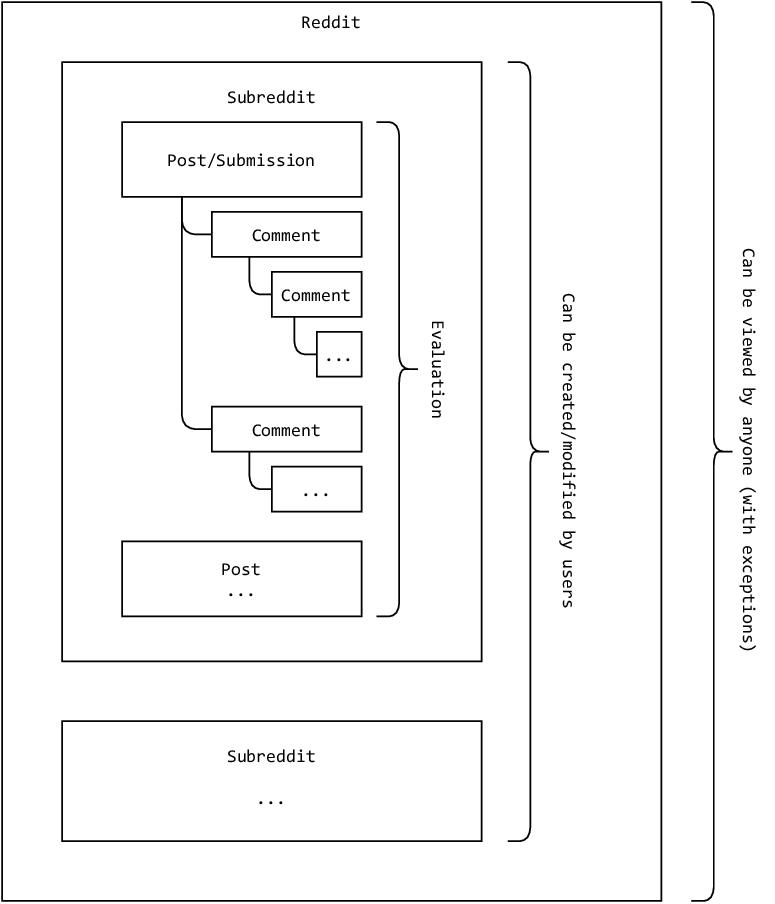}
    \caption{Reddit structure}
    \label{fig:reddit_structure}
\end{figure}

Let us start from a brief description of Reddit. It is a web content rating and discussion website~\cite{medvedev2017anatomy}. It was created in 2005 and is ranked as the 17\textsuperscript{th} most visited website in the world, with over 430 million monthly active users\footnote{\url{https://www.statista.com/topics/5672/reddit/\#dossierSummary}} and total of over 13 billion posts and comments\footnote{\url{https://www.redditinc.com/}}. The structure of Reddit is illustrated in Figure~\ref{fig:reddit_structure}.

Reddit is divided into thematic subfora (so called, \textit{subreddits}) dynamically created by its users. Therefore, the topic structure is systematically evolving, in response to user needs. Each subreddit has its \textit{moderators} who may supervise \textit{submissions} and \textit{comments}. Comments are linked to submissions, or to earlier comments, forming a tree-like structure.

\subsection{Content access rules and restrictions}

Most of the subreddits are public (for registered and non-registred users).
There are some exceptions based, for instance, on karma points (i.e. user's score), comments, gold (i.e. Reddit's currency that can be purchased with real money), moderator status, time on Reddit, username and others. For instance, such restriction can be be applied to even a Harry Potter house preference (e.g. r/gryffindor)\footnote{\url{https://www.reddit.com/r/ListOfSubreddits/wiki/privates}}. Here, let us note that the Reddit topic explorations tool (introduced in Section~\ref{sec:reddit-tudfe}), is based only on access to publicly available data.

\subsection{Accessibility -- Reddit API vs. Pushshift API}

Not only is the data on Reddit publicly accessible (with the exception of private communities), it is also made available via the official Reddit API\footnote{\url{https://www.reddit.com/dev/api/}}. However, in the course of literature review, it was found that most researchers do not actually use it. Over 90\% of analyzed papers either use ready datasets scraped earlier from Reddit and posted online (possibly in an annotated form), or they choose the Pushshift API~\cite{baumgartner2020pushshift}. None of the analysed papers stated the explicit reason for this choice (very few even mention how their datasets have been retrieved). However, practically testing capabilities of Reddit API and Pushshift API shows that the key factor could have been that Reddit API does not allow easy retrieval of historical data, while Pushshift API does. Hence, when developing the Reddit data exploration tool, the Pushshift API was used.

\section{Data acquisition and processing} \label{sec:data-acquisition}

To explore Reddit, as seen by the scientists, a dataset of all, most recent, papers available on arXiv has been assembled -- a total of 180 papers. All of them were related to Reddit and submitted to arXiv between 01-01-2019 and 01-03-2021 (and retrieved on 30-03-2021\footnote{\url{https://arxiv.org/search/advanced?\&terms-0-term=reddit\&classification-computer_science=y\&date-from_date=2019-01-01\&date-to_date=2021-03-01}}). This dataset has been processed both manually and automatically. First, collected papers have been manually annotated with four attribute sets: \textbf{topic} (a general area of research), \textbf{methods} (theoretical approach, e.g. neural network, text embedding), \textbf{dataset} and \textbf{technologies} (practical software, e.g. BERT~\cite{devlin2018bert}). Next, obtained results were merged using arXiv identification code and the publicly available data, i.e. the content (title and raw text) and the bibliometric metadata. This allowed extraction of information presented in Section~\ref{sec:analysis-and-findings}. All collected content has been converted to a raw text file, using PDF Miner software~\cite{shinyama2015pdfminer}. Next, the key features of titles and texts have been cleaned and mined using the NLTK framework~\cite{loper2002nltk} (for sentiment and subjectivity), and TF-IDF~\cite{rajaraman_ullman_2011} for vectorization (both frameworks are part of the scikit-klearn library~\cite{scikit-learn}).

\section{Analysis and findings} \label{sec:analysis-and-findings}

As a result of processing of collected data, we were able to formulate a number of observations. Let us summarize the most important ones.

\subsection{Metadata and bibliometrics}

\begin{figure*}[htp]
    \centering
    \includegraphics[width=\linewidth]{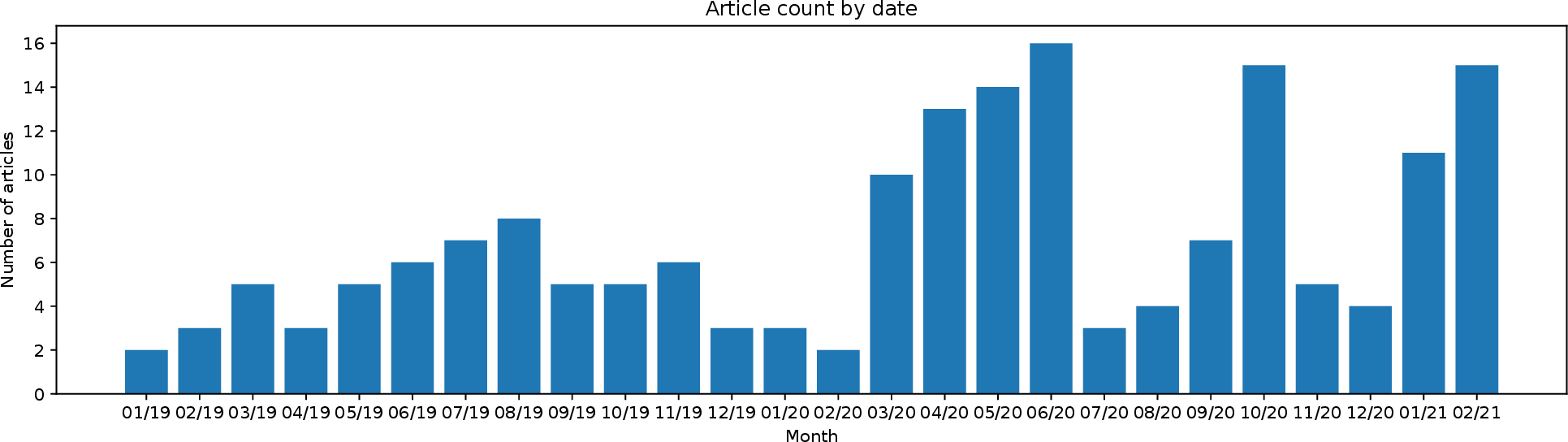}
    \caption{Article count by the month of the submission date}
    \label{fig:articles_month_histogram}
\end{figure*}

\begin{figure}[htpb]
    \centering
    \includegraphics[width=\linewidth]{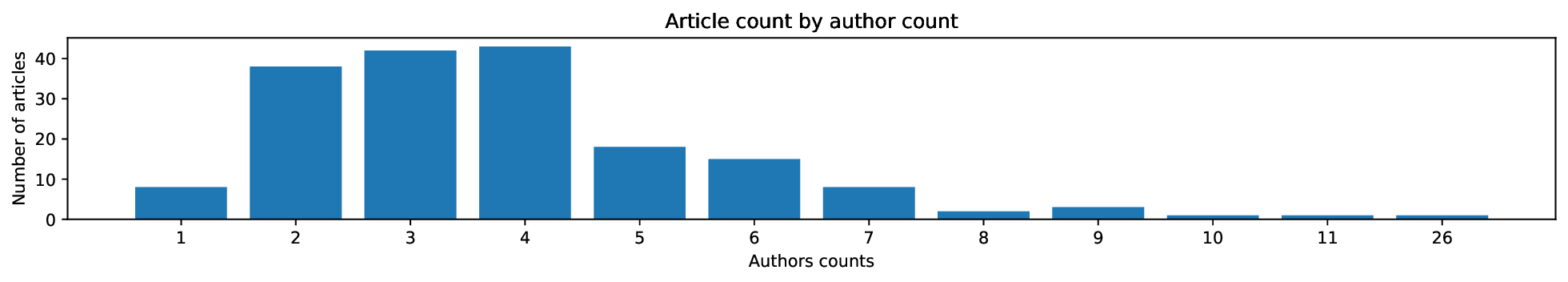}
    \caption{Number of authors for the selected papers}
    \label{fig:authors_counts_histogram}
\end{figure}

\begin{figure*}[htpb]
    \centering
    \includegraphics[width=\linewidth]{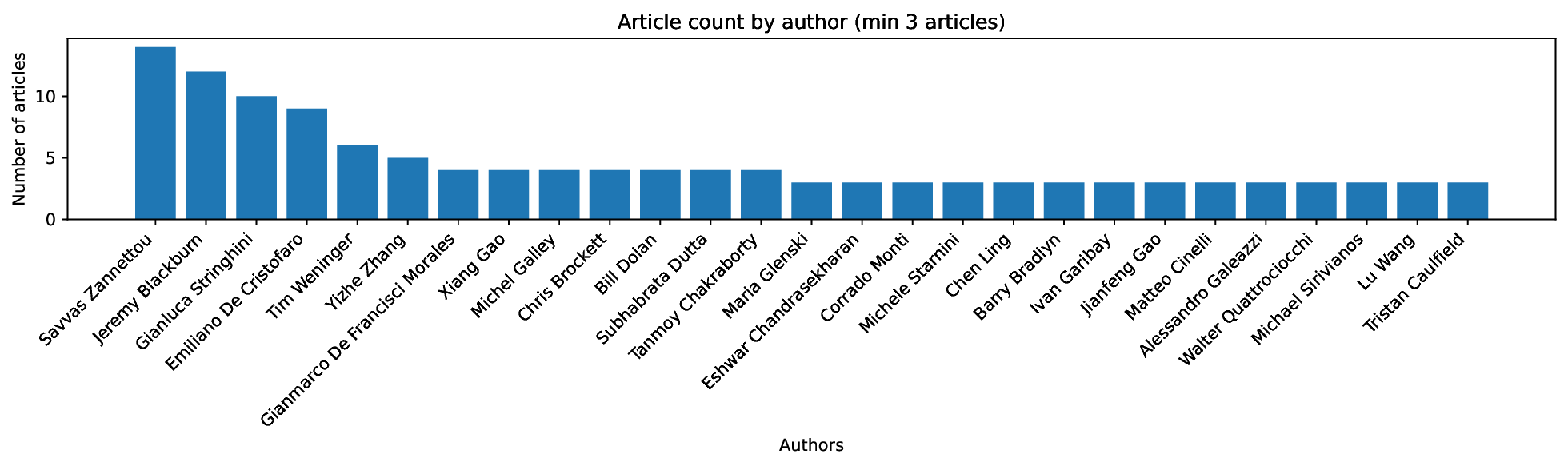}
    \caption{Article count by the author}
    \label{fig:authors_histogram}
\end{figure*}

First, let us consider a few noticeable bibliometric and authorship statistics, gathered using Semantic Scholar\footnote{\url{https://api.semanticscholar.org/}} and presented in Figures~\ref{fig:articles_month_histogram}, \ref{fig:authors_counts_histogram} and~\ref{fig:authors_histogram}. 

As shown in Figure~\ref{fig:articles_month_histogram}, there is a significant growth in the number of articles (related to Reddit) published after March 2020 (correlated with the outburst of the COVID-19 pandemic) and in October 2020 (correlated with notification dates for many scientific conferences~\cite{viglione2020scientific}). The latter fact was also verified during manual processing of collected data. This suggests that Reddit was used to provide data related to COVID pandemic and that it is used as a data source for contributions to, broadly understood, data analytics related conferences.

Next, as seen in Figure~\ref{fig:authors_counts_histogram}, majority of papers were written by 2-4 authors, with one having 26 authors~\cite{garibay2020deep}. 

Finally, Figure~\ref{fig:authors_histogram}, shows that the most prolific authors, of Reddit-related papers, were Savvas Zannettou (Max-Planck-Institute), Jeremy Blackburn (Binghamton University) and Gianluca Stringhini (Boston University). This seems to suggest that large number of scientific content, generated while studying Reddit posts, is delivered by a close circle of scientists.

\subsection{Analysis of topic, methods and technology}

\begin{figure*}[htpb]
    \centering
    \includegraphics[width=\linewidth]{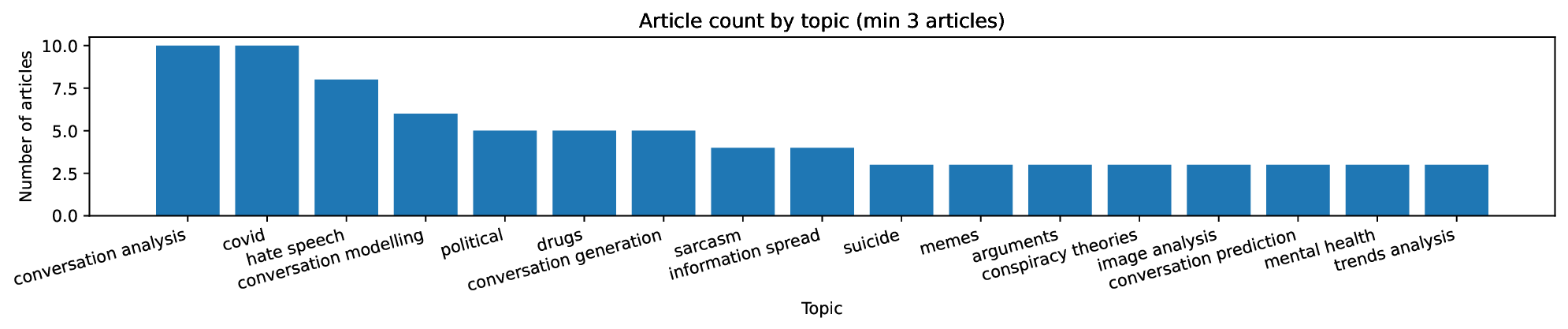}
    \caption{Article count by article topics manually annotated in all papers}
    \label{fig:topics_histogram}
\end{figure*}

\begin{figure*}[htpb]
   \centering
    \includegraphics[width=\linewidth]{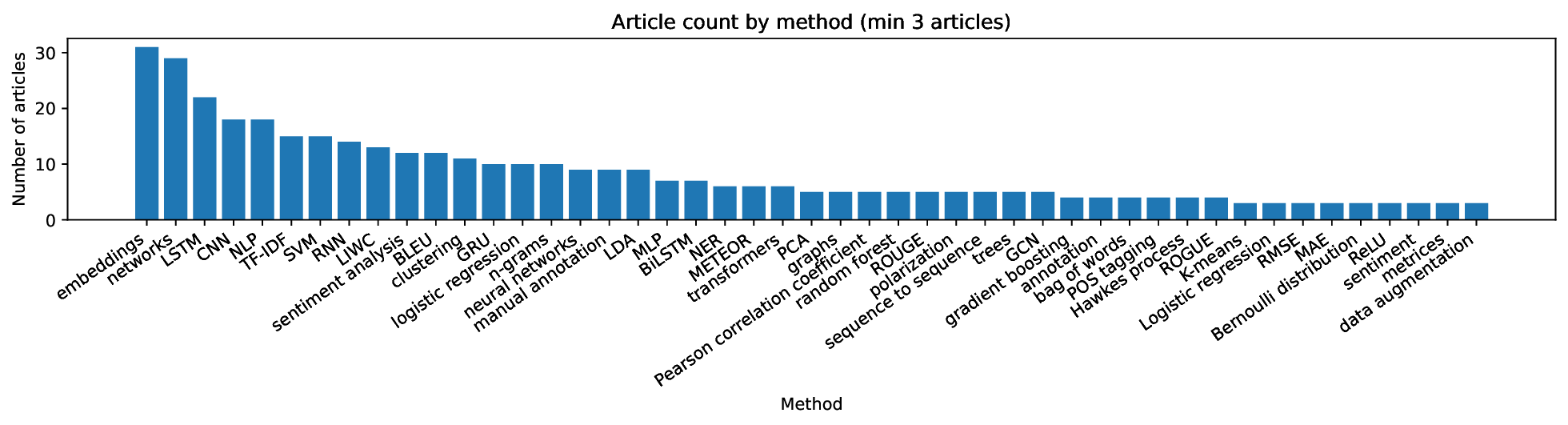}
   \caption{Article count by methods manually annotated in all papers}
    \label{fig:methods_histogram}
\end{figure*}

\begin{figure*}[htpb]
   \centering
    \includegraphics[width=0.8\linewidth]{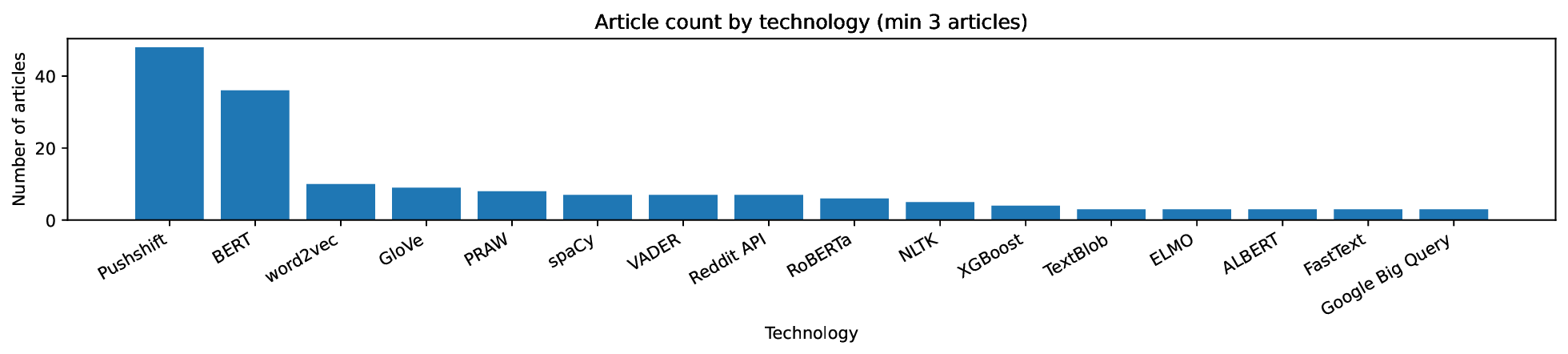}
   \caption{Article count by technologies manually annotated in all papers}
    \label{fig:technologies_histogram}
\end{figure*}

Topics, methods and technologies are key to answer RQ1 and RQ2.
These were extracted manually from the collected papers. 
They are summarized in Figures \ref{fig:topics_histogram}, \ref{fig:methods_histogram} and \ref{fig:technologies_histogram}. 

Figures~\ref{fig:topics_histogram} and~\ref{fig:methods_histogram} show clearly that the most popular research topic is \textit{conversation}, which matches the fact that Reddit is a discussion forum. Due to the timing of this work (overlapping with the COVID-19 pandemic), the second most common topic is \textit{COVID} (see Figure~\ref{fig:topics_histogram}).

Since Reddit consists mostly of text-based discussions, it is not surprising that the two most common methods, in Reddit-related research, are \textit{text embeddings}, used in text processing, and \textit{networks}, used for social network analysis. Note that, in the reported results, ``network'' (understood as a graph) and ``neural network'' are separate terms.

Regarding technologies (shown in Figure~\ref{fig:technologies_histogram}), over 45\% of studies used Pushshift API~\cite{baumgartner2020pushshift} for Reddit data extraction, and over 35\% applied BERT~\cite{devlin2018bert} embedding (and its variations) for the natural language processing.

\begin{figure*}[htpb]
    \centering
    \includegraphics[width=0.7\linewidth]{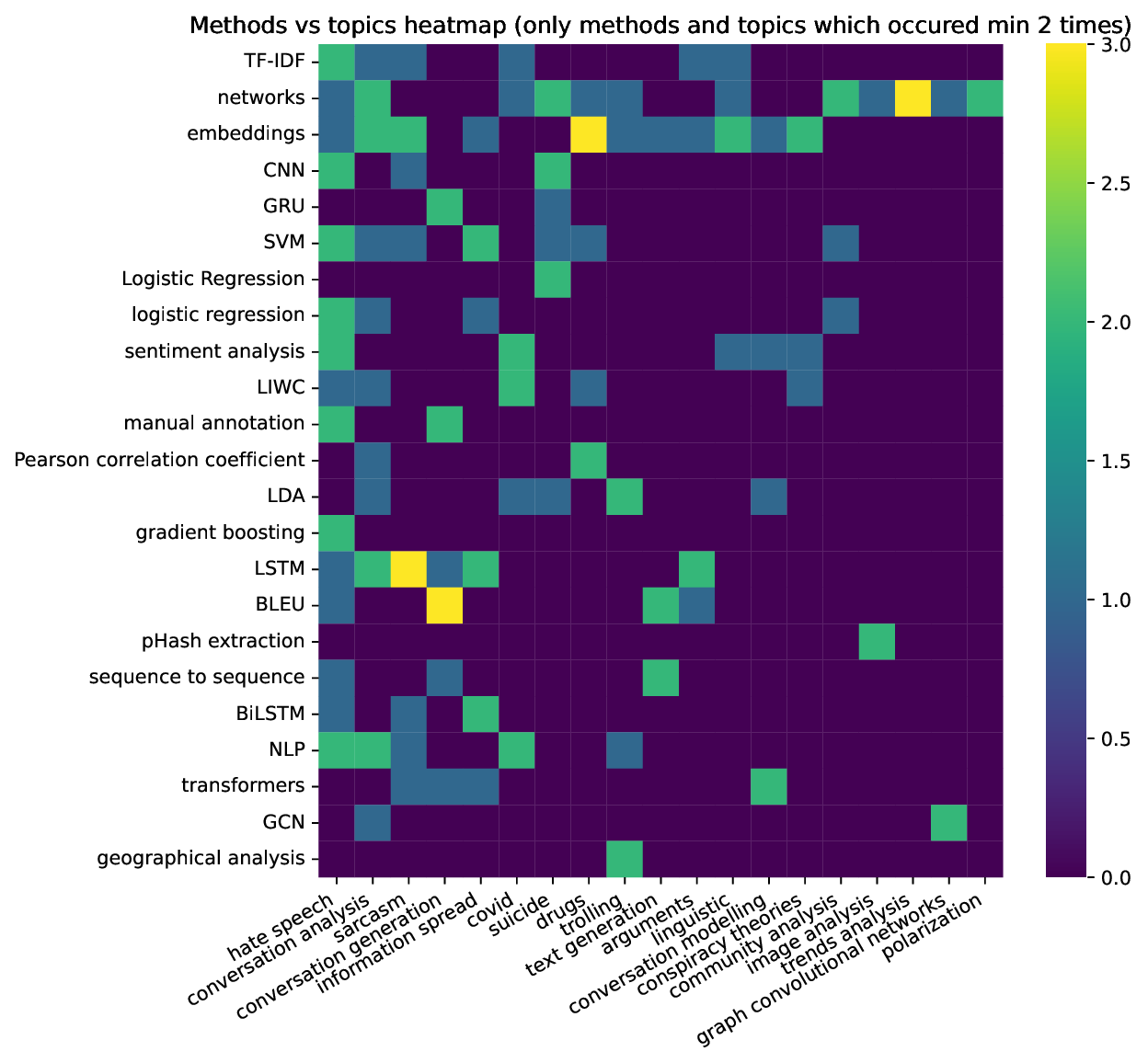}
    \caption{Research methods correlated with article topics}
    \label{fig:topics_methods_heatmap}
\end{figure*}

Finally, topics and methods have been combined in a correlation heatmap (Figure~\ref{fig:topics_methods_heatmap}). 

Here, a few significant correlations have been established. However, they have to be considered keeping in mind that they materialize in the context of a specific dataset, created on from contributions reporting research that used Reddit as a data source. therefore, no claim is made that these observation can be immediately generalized beyond the dataset used in this work. However, based on general knowledge of the field, they seem to be in line with more general trends.
\begin{itemize}
    \item Papers related to \textit{drugs} typically use \textit{word embeddings}. However, this can be related to the overall popularity of word embeddings in the research conducted in early 2020th (see, for instance, the citation count for~\cite{devlin2018bert}). 

    \item \textit{Networks} are typically applied in analysis of \textit{trends}, e.g. topic popularity (this is a key finding for RQ1).
    
    \item Articles dealing with \textit{sarcasm} often use \textit{LSTM networks}.
    
    \item Research devoted to the \textit{conversation generation} typically applies the \textit{BLEU metric}. 
\end{itemize}

\subsubsection{Topics of knowledge and information processing}

The topic of information and knowledge retrieval is one of the main aims of undertaken analysis. Hence, this category was checked specifically. Even though many works focus on information spreading in online communities~\cite{garibay2020deep,zannettou2019towards,edelbo2019danish,fajcik2019but}, there is hardly any focus purely on information/know\-ledge retrieval.
There are precisely two papers (1\% of the considered work) related to knowledge processing (specifically, knowledge graphs~\cite{cao2020building,zhang2019grounded}). Expanding arXiv search, to capture all articles including terms ``knowledge'' and ``Reddit'', resulted in 4 records, none of which is related to knowledge capture. Pairing keyword ``Reddit'' with ``information retrieval'' or 'information processing'' yielded 0 results. Therefore, top knowledge processing/management-related conferences were searched, but only one contribution~\cite{hastings2011model}, about knowledge and Reddit, has been found (published by the K-CAP conference in 2011\footnote{\url{https://www.k-cap.org/kcap11/index.html}}). This renders Reddit as a source that is definitely underexplored in terms of knowledge/information mining.

\subsubsection{Use of Reddit combined with other datasets}

\begin{figure}[htpb]
    \centering
    \includegraphics[width=\linewidth]{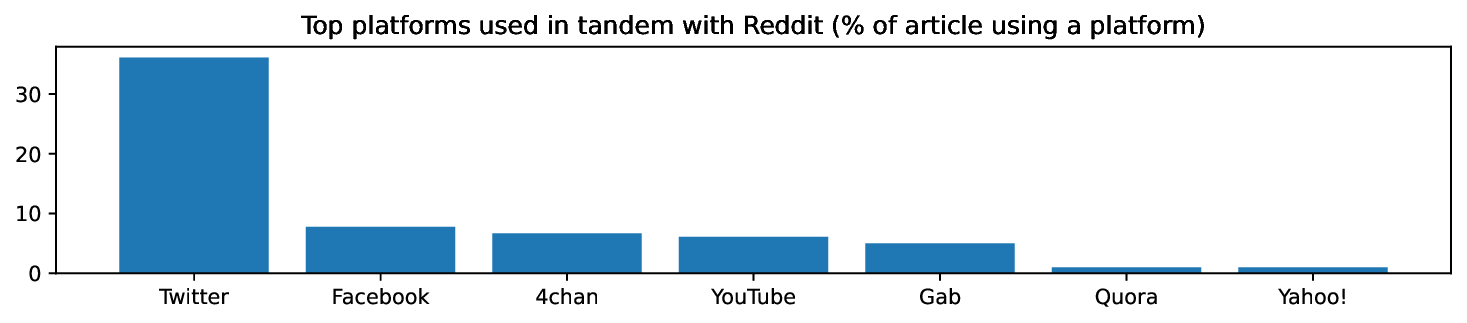}
    \caption{Online platforms used as data sources together with Reddit}
    \label{fig:platform_dataset_count}
\end{figure}

Moving to the \textbf{RQ5}, it was discovered that among papers that use Reddit, over 30\% also use Twitter, which is a data source that is very often used for sentiment analysis~\cite{kharde2016sentiment}). Other datasets that have been utilized together with Reddit are: Facebook, 4Chan, YouTube, and Gab. 
Each of them appears in less than 10\% of papers, which used Reddit (details are shown in Figure \ref{fig:platform_dataset_count}). Datasets are rarely used in triplets, i.e. Reddit and two other datasets (the highest scoring triplets were Reddit, combined with Twitter and Facebook ~6.6\% of articles; Reddit, used together with Twitter and 4chan ~6\% (e.g.~\cite{zannettou2019towards,zannettou2019disinformation}), and Reddit studied jointly with Twitter, YouTube ~5\% of contributions (e.g.~\cite{buntain2021youtube})). Finally, a single paper considers combination of four datasets (i.e. Reddit, Twitter, Facebook, and Gab~\cite{cinelli2020echo}).

An interesting use case of Reddit usage in scientific environment has been found in ``IEEE Top Programming Languages: Design, Methods, and Data Sources''\footnote{\url{https://spectrum.ieee.org/ieee-top-programming-languages-design-methods-and-data-sources}}. 
This work shows a practical approach to a an interesting research question; here, what are the top programming languages. 
In this work Reddit is listed as one of the sources among others, such as Google Trends, Twitter, GitHub and Stack Overflow. 

\subsection{Linguistic analysis}\label{sec:linguistic-analysis}

\begin{figure}[htpb]
    \centering
    \includegraphics[width=\linewidth]{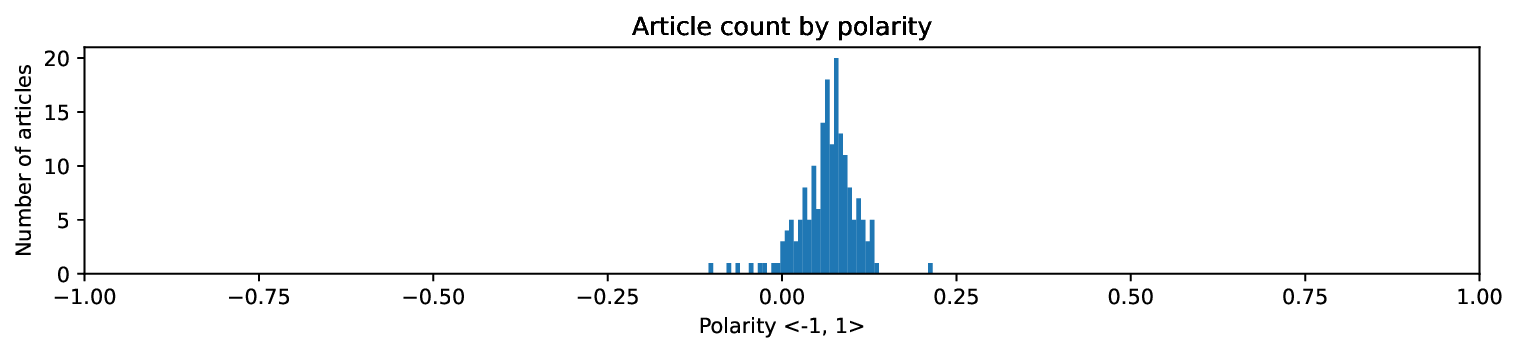}
    \caption{Histogram of number articles based on text polarity measures}
    \label{fig:polarity_histogram}
\end{figure}

During exploratory data analysis, various natural language processing techniques were applied. 
Among them, papers were also analysed linguistically. Specifically, sentiment analysis using NLTK framework~\cite{loper2002nltk} and SentimentAnalyzer\footnote{\url{https://www.nltk.org/api/nltk.sentiment.sentiment_analyzer.html}} was applied. 
Observed polarization (depicted in Figure~\ref{fig:polarity_histogram}) indicates a negligible displacement towards the positive sentiment. This was expected, and is consistent with previous studies on scientific literature sentiment~\cite{hussein2016analyzing}. 

\begin{figure}[htpb]
    \centering
    \includegraphics[width=\linewidth]{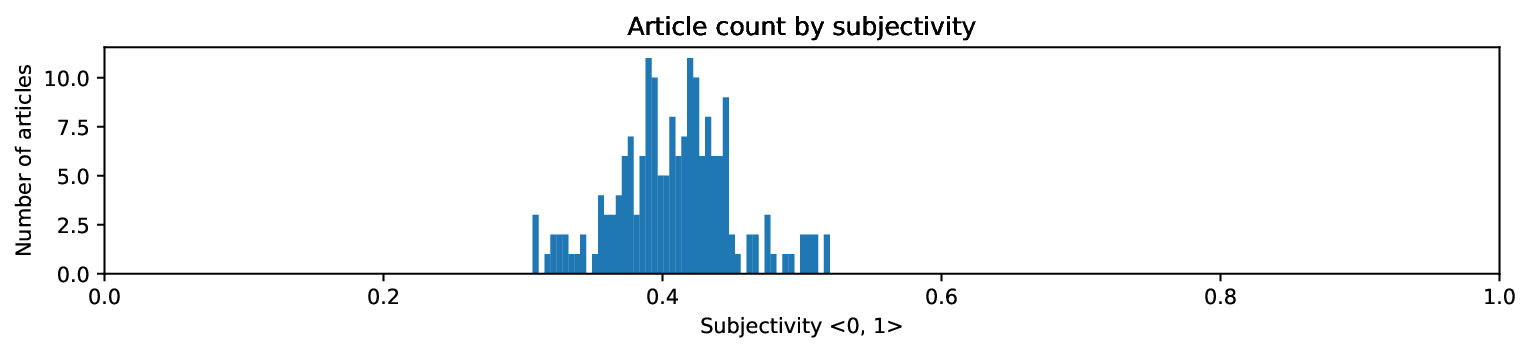}
    \caption{Histogram of number articles based on text subjectivity measures}
    \label{fig:subjectivity_histogram}
\end{figure}

However, the subjectivity measure (summarised in Figure~\ref{fig:subjectivity_histogram}) raised concerns. Obviously, it has been claimed that scientific research may be subjective, as it needs to allow ``leaps of faith'' (see,~\cite{curtis2012science}). Moreover, some philosophers~\cite{mannan2016science,mackellar2012subjectivity} argue that subjectivity is intrinsic for human nature. However, it is also claimed (and for good reasons) that the foundation of the scientific method~\cite{newton1833philosophiae} revolves around aiming at objectivity. Hence, results summarised in Figure~\ref{fig:subjectivity_histogram}, indicating high level of subjectivity, were somewhat concerning. To establish the reason for this finding, the most ``subjective'' texts were studied directly. As a result is was found that this is a false alarm. Specifically, apparent shift towards subjectivity was caused by inaccuracy of the classifier (\textit{SentimentIntensityAnalyzer} from \textit{nltk.sentiment}\footnote{\url{https://www.nltk.org/api/nltk.sentiment.html}}). For further understanding, let us consider the selected sentences from the most subjective (according to the NLTK metric) articles.

\begin{itemize}
    \item ``However, this openness formed a platform for the polarization of opinions and controversial discussions''~\cite{jasser2020controversial} (score: 0.95)
    \item ``(...) also presented an extended version of the study discussing potential racial bias in offensive content datasets (...)''~\cite{aggarwal2020trawling} (score: 1.0)
    \item ``All datasets only contain activity between 01/2015 and 10/2018''~\cite{habib2019act} (score: 1.0)
\end{itemize}

Moreover, let us also consider how the calculated subjectivity measure changes with a simple modification of selected statements (i.e. by removing particular words): 

\begin{itemize}
    \item Statement before transformation (score: 0.63): 
    
    ``Controversially initiated and non-controversially initiated cascades, (a,b,c) are controversially initiated posts’ cascades while (d,e,f) are non-controversial posts’ cascades where the red dots represent a comment labeled as controversial by Reddit that is directed to the post’s author while a green dot is a comment labeled controversial by Reddit that is directed to another comment.''~\cite{jasser2020controversial}
    
    \item The same statement after transformation (score: 0.15): 
    
    ``initiated and initiated cascades, (a,b,c) are initiated posts’ cascades while (d,e,f) are posts’ cascades where the red dots represent a comment labeled as by Reddit that is directed to the post’s author while a green dot is a comment labeled by Reddit that is directed to another comment.''~\cite{jasser2020controversial}
\end{itemize}

This suggests that simply using the ``subjective'' (key)words (e.g. ``controversial'', ``bias'') in the text, regardless of their context, results in radically increased value of the variable that is to indicate subjectivity of the text. However, there are sentences that do not use such words, which have also received a high subjectivity score. Hence, further research would be required into the way that the NLTK metric works and why, sometimes, it is rather misleading. However, this is outside of scope of the current contribution.

\subsection{Reddit-based literature in scholarly databases}

\begin{figure*}[htpb]
    \centering
    \includegraphics[width=\linewidth]{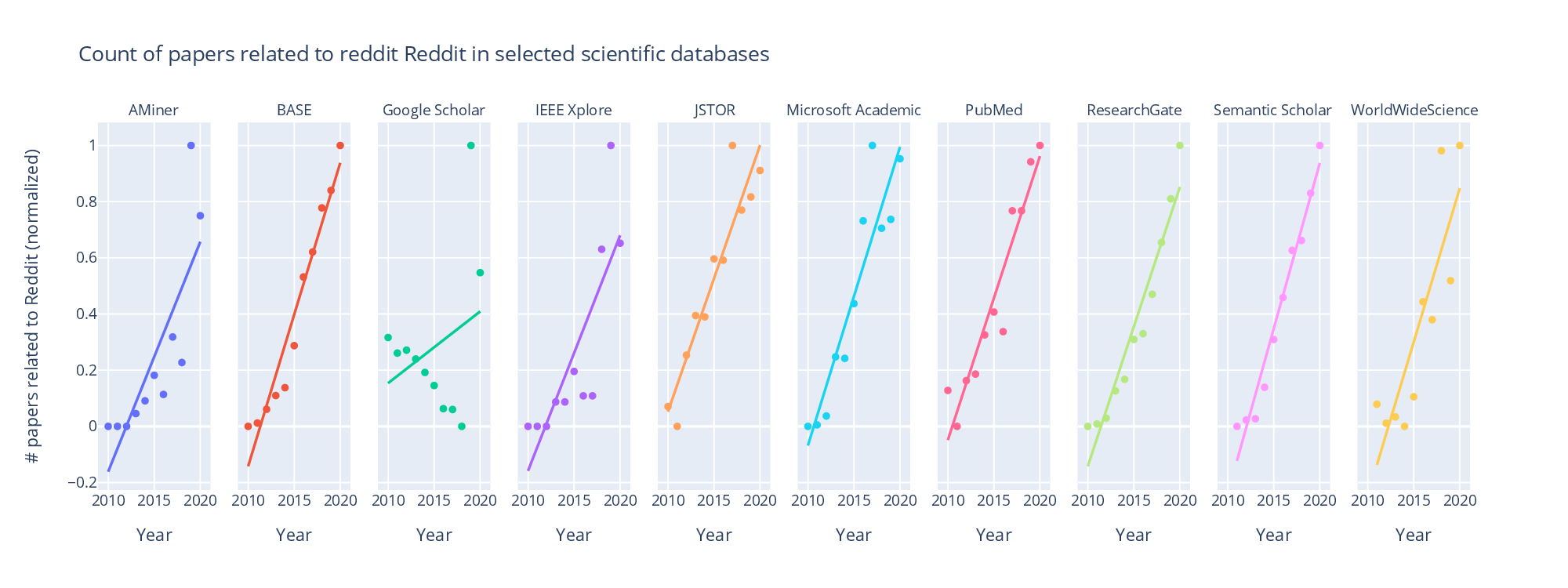}
    \caption{Non-cumulative count of papers related to Reddit in scientific databases in years 2010-2021.}
    \label{fig:reddit_scientific_databases}
\end{figure*}

Let us now address \textbf{RQ3} and \textbf{RQ4}. Even though they cannot be unequivocally answered, possible answers can be experimentally explored. To verify the change over time of the number of scholarly papers related to Reddit, between 2010 and 2021, 10 databases have been analysed and queried for the term ``reddit''. As shown in Figure~\ref{fig:reddit_scientific_databases} the number of found articles raises year to year (\textbf{RQ3}). 

Table \ref{tab:reddit-scientific-paper-databases} shows how many articles, related to Reddit (e.g. using it as a data source, processing it, analysing the comments, etc.), have been indexed in scientific databases. 

\begin{table}[htpb]
  \caption{Number of scientific papers (which appeared in 2011-2021) related to Reddit (e.g. using it as dataset, exploring its structure etc.) indexed in selected databases (all accessed on 09-06-2021).}
  \label{tab:commands}
  \begin{tabular}{ccl}
    \toprule
    \begin{tabular}{p{5cm}|p{4cm}|p{2cm}}
Database &
  Total size &
  \# papers \\
\midrule
Google Scholar &
  330M~\cite{gusenbauer2019google} &
  980\footnote{\url{https://scholar.google.com/scholar?start=980\&q=reddit\&as\_ylo=2010}} \\
JSTOR &
  12M\footnote{\url{https://about.jstor.org/}} &
  2555\footnote{https://www.jstor.org/action/doBasicSearch?Query=reddit\&sd=2011\&ed=2021} \\
PubMed &
  32M\footnote{\url{https://pubmed.ncbi.nlm.nih.gov/}} &
  243\footnote{\url{https://pubmed.ncbi.nlm.nih.gov/?term=reddit\&filter=dates.2010-2021}} \\
AMiner &
  230M~\cite{gusenbauer2019google} &
  840\footnote{\url{https://www.aminer.org/search/pub?q=reddit\&t=b\&time=2010-2021}} \\
Bielefeld Academic Search Engine &
  270M\footnote{\url{https://www.base-search.net/}} &
  5176\footnote{\url{https://www.base-search.net/Search/Results?lookfor=reddit\&type=all\&l=pl\&oaboost=1\&refid=dcfilpl} }\\
Semantic Scholar &
  197M\footnote{\url{https://www.semanticscholar.org/}} &
  2280\footnote{\url{https://www.semanticscholar.org/search?year\%5B0\%5D=2010\&year\%5B1\%5D=2021\&q=reddit\&sort=relevance}} \\
AMiner & 
  320M\footnote{\url{https://www.aminer.org/}} &
  784\footnote{\url{https://www.aminer.org/search/pub?q=reddit\&t=b\&time=2010-2021}} \\
Microsoft Academic & 
170M~\cite{gusenbauer2019google} &
  902\footnote{\url{https://academic.microsoft.com/search?q=reddit\&f=\&eyl=Y\%3C\%3D2021\&syl=Y\%3E\%3D2010\&orderBy=0}} \\
WorldWideScience &
  300M~\cite{gusenbauer2019google} &
  1250\footnote{\url{https://worldwidescience.org/wws/desktop/en/results.html}} \\
IEEE Xplore &
  8551\footnote{\url{https://ieeexplore.ieee.org/browse/conferences/title?ranges=2011_2021_Year}} &
  162\footnote{\url{https://ieeexplore.ieee.org/search/searchresult.jsp?queryText=reddit\&highlight=true\&returnFacets=ALL\&returnType=SEARCH\&matchPubs=true\&ranges=2010\_2021\_Year}}\\
ResearchGate &
  135M\footnote{\url{https://www.researchgate.net/about}} &
  3001\footnote{\url{https://www.researchgate.net/search.Search.html?type=publication\&query=reddit}} \\
\bottomrule
\end{tabular}
\label{tab:reddit-scientific-paper-databases}
\end{tabular}
\end{table}

Observations that can be made, on the basis of the results found in  Table~\ref{tab:reddit-scientific-paper-databases}, are: 
\begin{itemize}
    \item the number of Reddit articles is quite small, yet representative,
    \item the number of Reddit papers is somewhat proportional in each database, so it can be stated that the literature is quite equally spread in the Internet.
\end{itemize}

\subsubsection{Outlying results found in Google Scholar}

The only database with trends inconsistent with others is Google Scholar (Figure~\ref{fig:reddit_scientific_databases}). However, although it is one of the most widely known databases that indexes scientific publications~\cite{lopez2019google,halevi2017suitability}, it already received both praise and  criticism~\cite{jacso2005google,shultz2007comparing}. 
Main problems of Google Scholar, pointed out in the literature, are: (i) difficulties to estimate the actual size of the database~\cite{orduna2015methods,jacso2008google}, (2) gender- and race-related bias in displaying contributions~\cite{jensenius2018benefits}, (iii) favoring incremental work~\cite{jensenius2018benefits}, (iv) favoring larger research communities~\cite{jensenius2018benefits}, (v) limited indexing of files~\cite{jacso2005google}, (vi) incorrect bibliometrics (due to automated algorithms instead of skilled librarians)~\cite{lopez2017google,gray2012scholarish}, (vi) ``uncertain quality of Google Scholar's performance''~\cite{gray2012scholarish}, (vii) ``Google Scholar's inability or unwillingness to elaborate on what documents its system crawls''~\cite{gray2012scholarish}, and (viii) limitations of bibliometric analysis~\cite{lopez2017google}. Moreover, Google Scholar declares inconsistently the number of results of a query, and the actual number of returned results (e.g. a query returns 1000 actual results,  while it declares 58,600\footnote{\url{https://scholar.google.com/scholar?start=990\&q=reddit\&hl=en\&as\_sdt=0,5\&as_ylo=2020\&as_yhi=2020} accessed on 11-09-2021}). This finding may correspond to already reported Google Scholar inconsistencies~\cite{orduna2015methods,jacso2008google} and lack of transparency~\cite{gray2012scholarish}. 
Therefore, Google Scholar can be treated as an outlier and disregarded in conclusions drawn from this experiment.

\subsection{Google Trends}

The next experiment explored presence of popular trends in Reddit. This was done based on Google Trends, an analytical website which provides information about popularity of search queries in Google search engine~\footnote{\url{https://trends.google.com/trends}}. For all Global Google Trends 2020\footnote{\url{https://trends.google.com/trends/yis/2020/GLOBAL/}} their Reddit presence has been measured (see Table~\ref{tab:google-trends-reddit}). Overall, 79\% of top Google Trends have a dedicated subreddit, while \textit{all of them} are widely discussed. Table~\ref{tab:google-trends-reddit} illustrates top three in each Google Trend category. 

\begin{table*}[htpb]
  \caption[aaaaaaaaaa]{Global Google Trends 2020\footnote{\url{https://trends.google.com/trends/yis/2020/GLOBAL/}} (top 3 in each Google Trends category) and their appearance on Reddit (``subreddit'' -- there exists a dedicated subforum, ``discussion'' -- the topic is present in (a) subreddit(s) of a broader topic)}
  \label{tab:google-trends-reddit}
    \begin{tabular}{p{3.5cm}|p{2cm}|p{2cm}|p{3.8cm}}
    \toprule
    Google Trend                 & category   & on Reddit         & reference               \\
    \midrule
    Coronavirus                  & searches   & subreddit         & r/Coronavirus           \\
    Election results             & searches   & discussion        & r/politics              \\
    Kobe Bryant                  & searches   & subreddit         & r/kobebryant            \\
    Tom Hanks                    & actors     & subreddit         & r/tomhanks              \\
    Joaquin Phoenix              & actors     & subreddit         & r/joaquinphoenix        \\
    Amitabh Bachchan             & actors     & subreddit         & r/india                 \\
    Ryan Newman                  & athletes   & subreddit         & r/RyanNewman            \\
    Michael Jordan               & athletes   & subreddit         & r/michaeljordan         \\
    Tyson Fury                   & athletes   & subreddit         & r/TysonFury             \\
    Parasite                     & movies     & subreddit         & r/parasite              \\
    1917                         & movies     & subreddit         & r/1917                  \\
    Black Panther                & movies     & subreddit         & r/blackpanther          \\
    Tiger King                   & tv shows   & subreddit         & r/TigerKing             \\
    Big Brother Brasil           & tv shows   & subreddit         & r/BigBrotherBrasil      \\
    Money Heist                  & tv shows   & subreddit         & r/MoneyHeist            \\
    Joe Biden                    & people     & subreddit         & r/JoeBiden              \\
    Kim Jong Un                  & people     & subreddit         & r/kimjongun             \\
    Boris Johnson                & people     & subreddit         & r/BorisJohnson          \\
    Coronavirus                  & news       & subreddit         & r/Coronavirus           \\
    Election results             & news       & discussion        & r/politics              \\
    Iran                         & news       & subreddit         & r/iran                  \\
    Among Us                     & games      & subreddit         & r/AmongUs               \\
    Fall Guys: Ultimate Knockout & games      & subreddit         & r/FallGuysGame          \\
    Valorant                     & games      & subreddit         & r/VALORANT              \\
    Dalgona coffee               & recipes    & discussion        & r/cafe                  \\
    Ekmek                        & recipes    & discussion        & r/Breadit               \\
    Sourdough bread              & recipes    & subreddit         & r/SourdoughBread        \\
    Kobe Bryant                  & loss       & subreddit         & r/kobebryant            \\
    Naya Rivera                  & loss       & subreddit         & r/NayaRivera            \\
    Chadwick Boseman             & loss       & subreddit         & r/ChadwickBoseman       \\
    \bottomrule
  \end{tabular}
\end{table*}

\section{Reddit as ``The Ultimate Dataset for Everything''} \label{sec:reddit-tudfe}

To further study whether Reddit contains information about (almost) ``any area'', a tool for easy exploratory data analysis (EDA~\cite{cox2017exploratory}) was designed. Specifically, 
\textit{Reddit-TUDFE} allows quick search of any topic on Reddit, checking if/how it is represented, and how it is discussed. Specifically, \textit{Reddit-TUDFE} delivers the following functions:

\begin{enumerate}
    \item Uses Reddit API to search for best matching subreddit. 
    \item Downloads newest $N$ posts from the subreddit, using Pushshift API and a combination of PRAW\footnote{\url{https://github.com/praw-dev/praw}} and PSAW\footnote{\url{https://github.com/dmarx/psaw}}.
    \item Performs basic text cleaning (tokenization with NLTK~\cite{loper2002nltk}, removal of stopwords, punctuation, numbers).
    \item Generates and displays post titles and content wordclouds\footnote{\url{https://github.com/amueller/word\_cloud}}.
\end{enumerate}

The code follows state-of-the-art solutions for code sharing (\cite{perkel2018jupyter}) and is publicly available on GitHub\footnote{\url{https://github.com/JanSawicki/reddit-tudfe/blob/main/reddit_tudfe.ipynb}} as a Jupyter Notebook~\cite{randles2017using}. 

\begin{figure}[htpb]
  \centering
    \begin{tabular}{c}
        \includegraphics[width=0.30\linewidth]{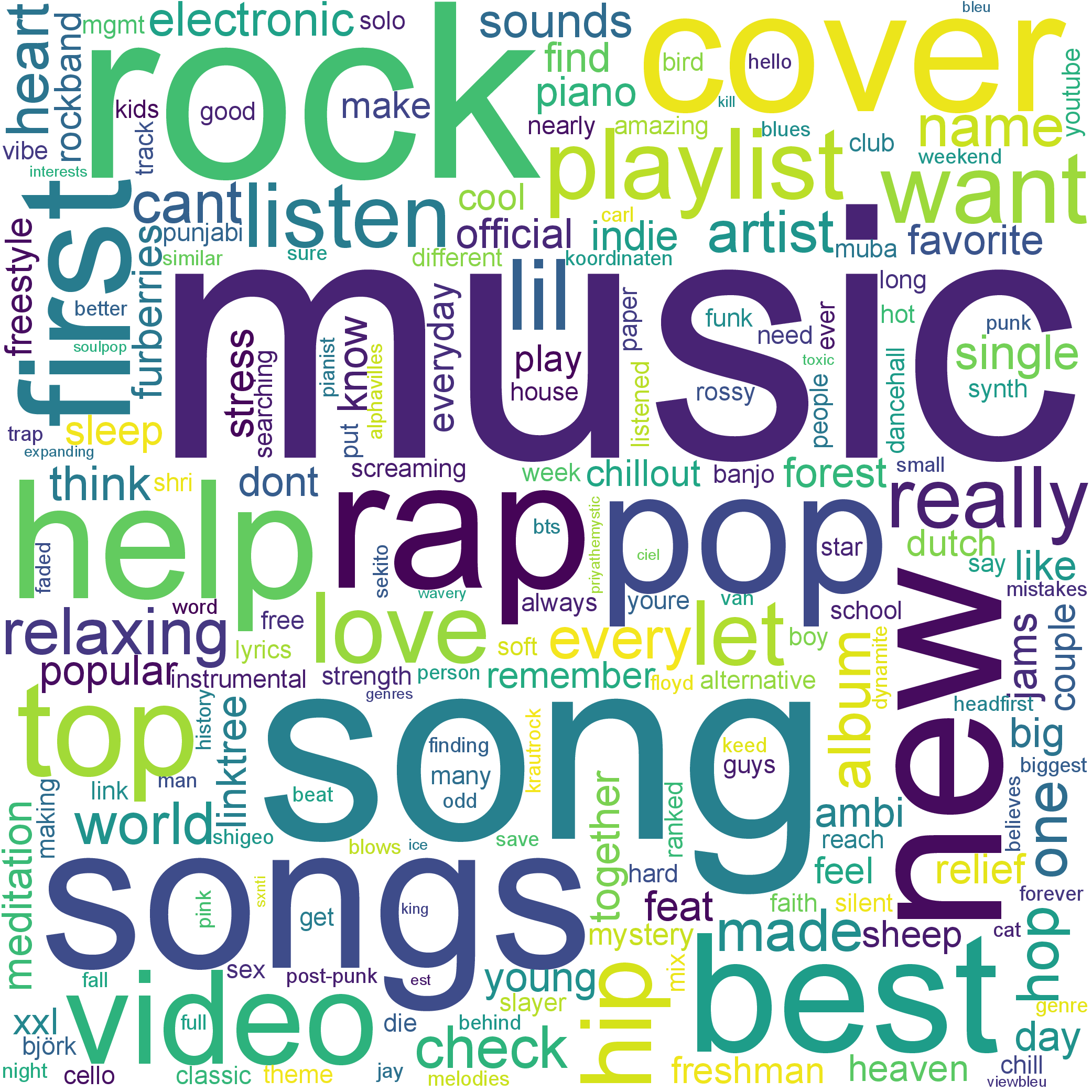}
        \includegraphics[width=0.30\linewidth]{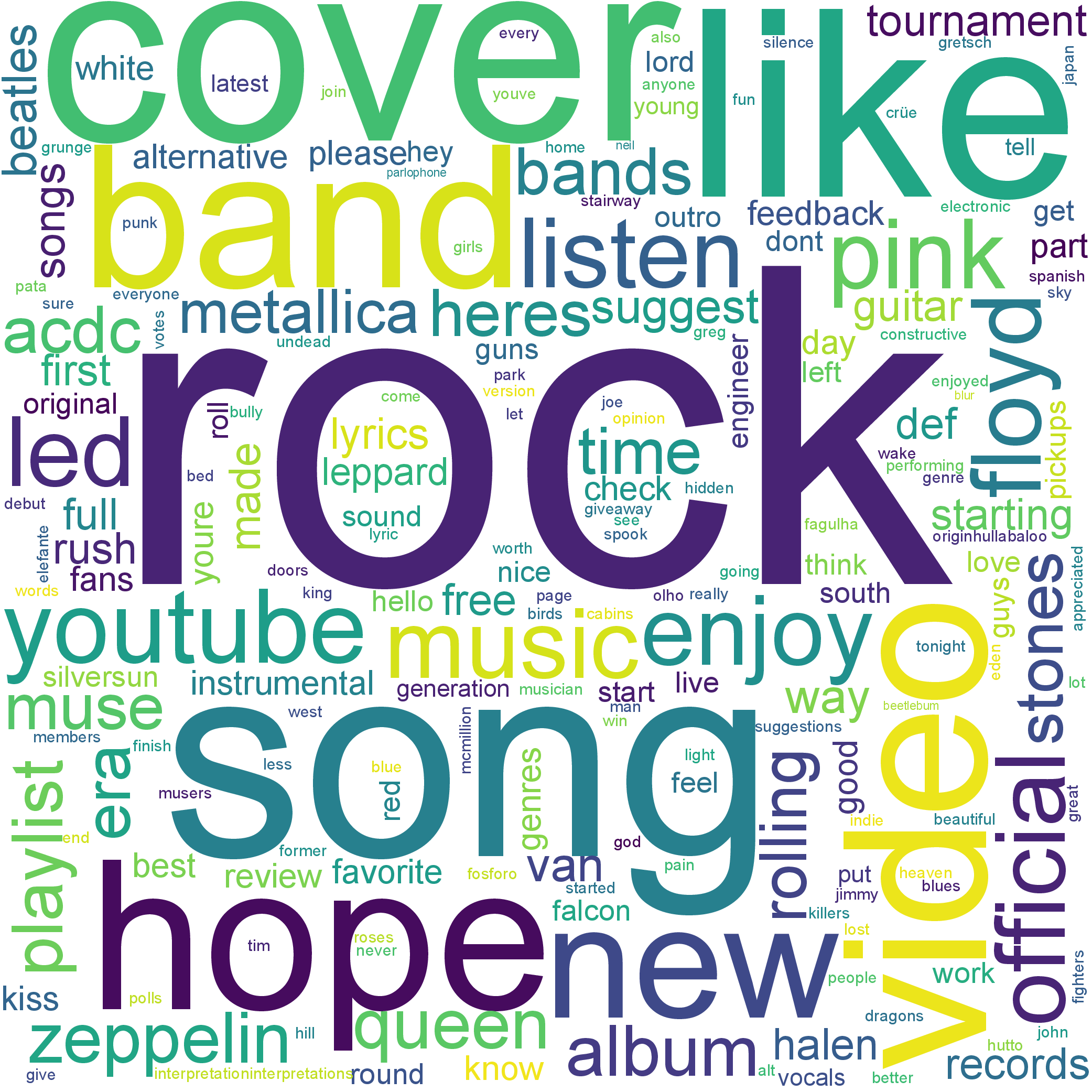}
        \includegraphics[width=0.30\linewidth]{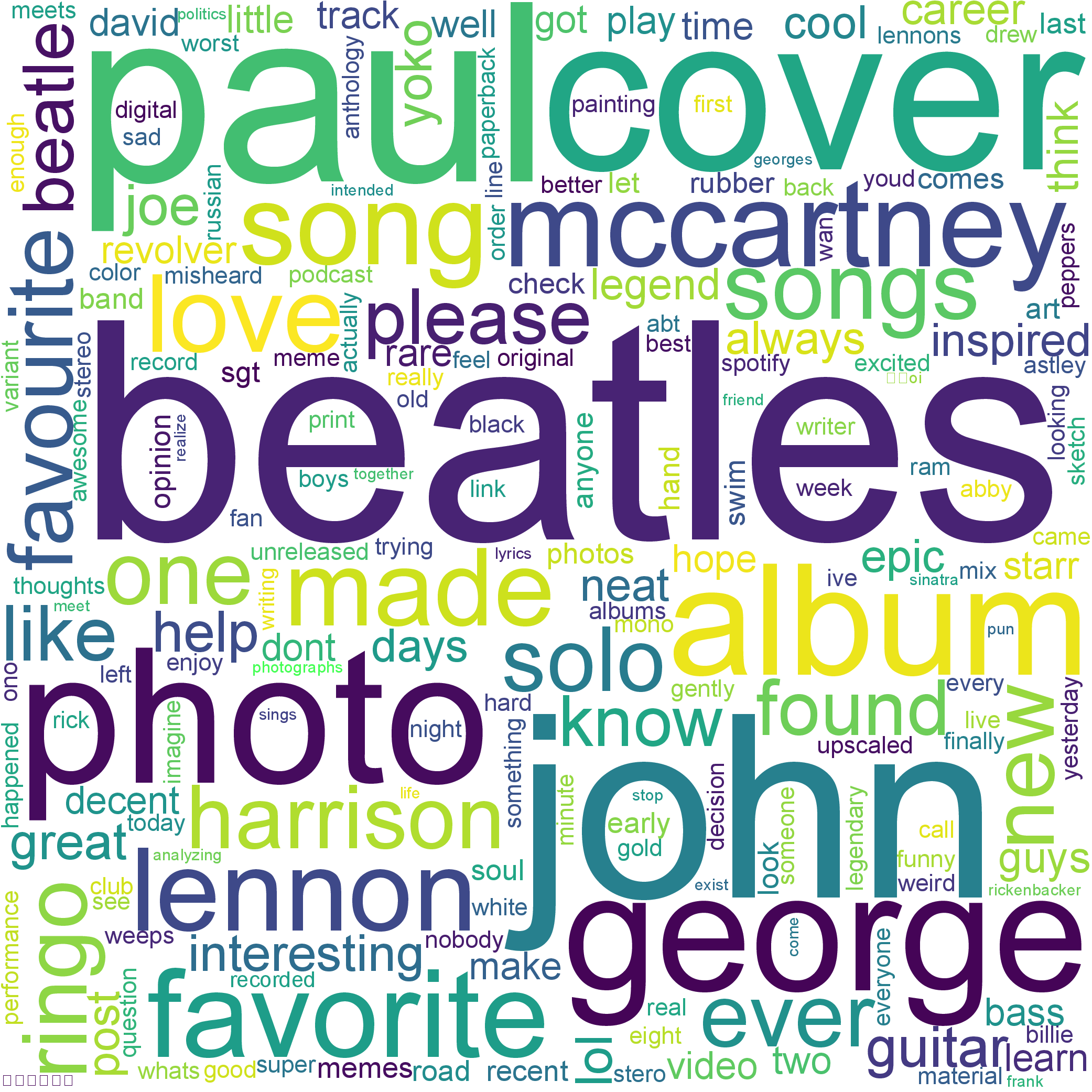}
    \end{tabular}
    \caption{Exemplary wordclouds of 200 posts (before 01-09-2021) concerning (top to bottom): ``music'', ``rock'' and ``The Beatles''.}
    \label{fig:reddit-tudfe-music}
\end{figure}

\begin{figure}[htpb]
  \centering
    \begin{tabular}{c}
        \includegraphics[width=0.30\linewidth]{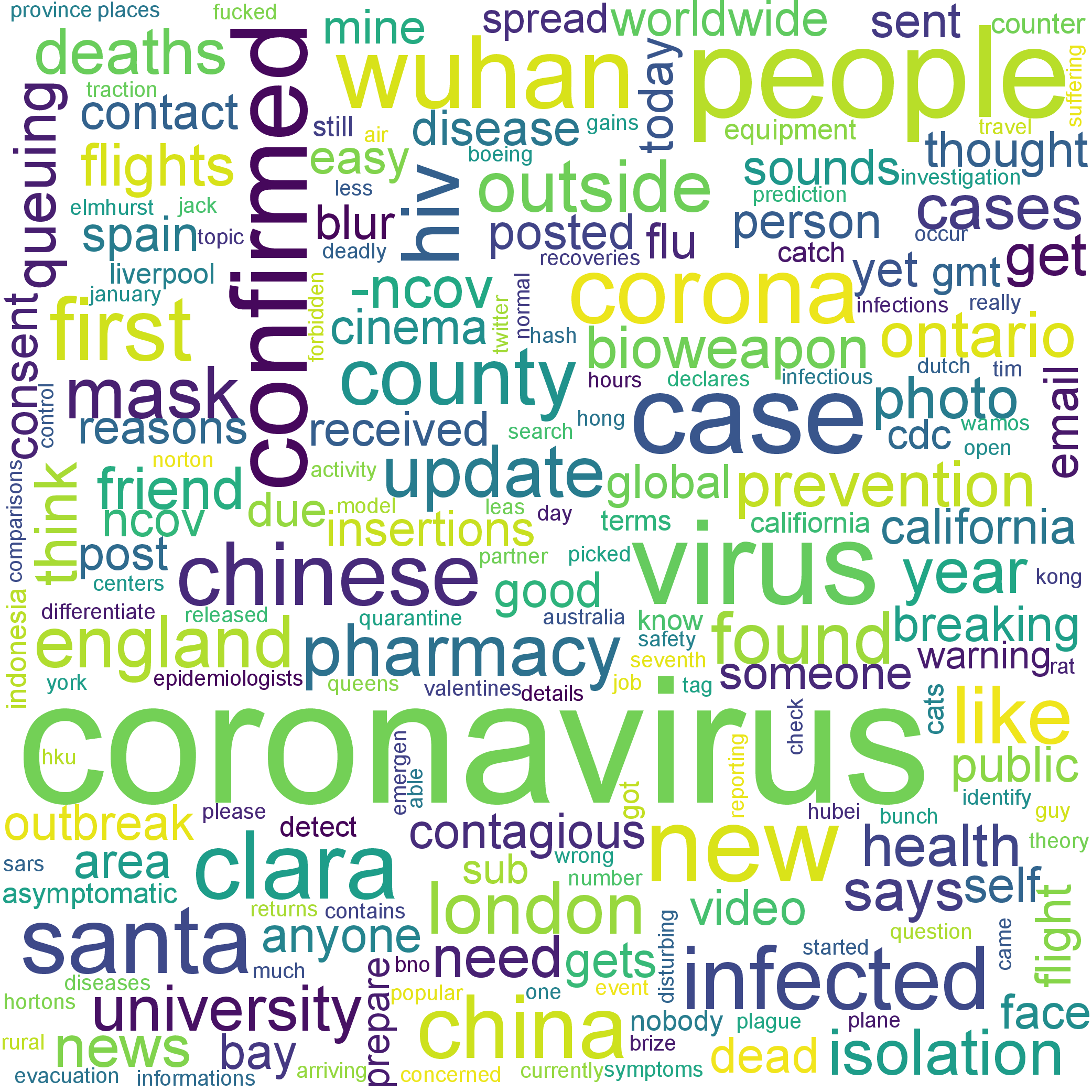}
        \includegraphics[width=0.30\linewidth]{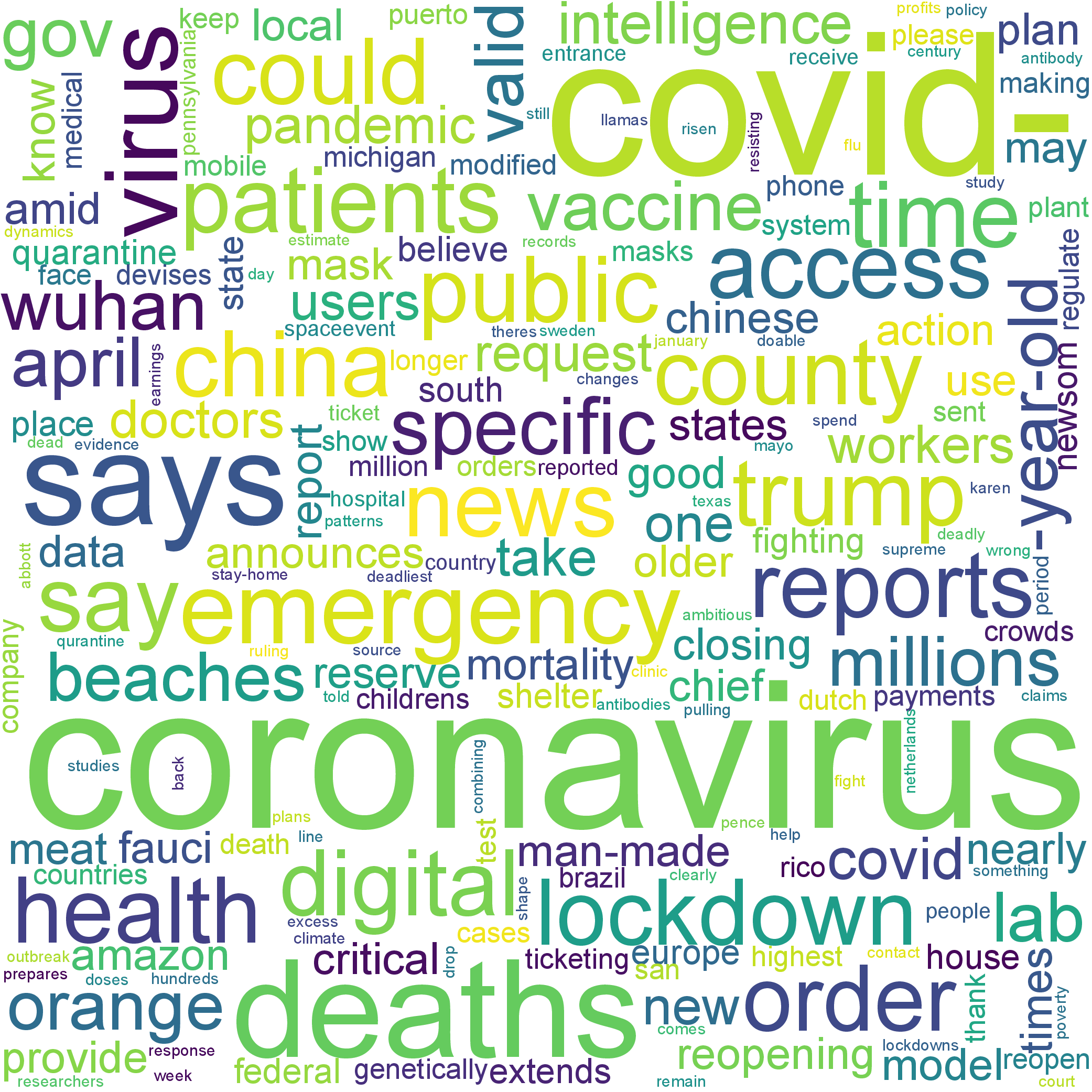}
        \includegraphics[width=0.30\linewidth]{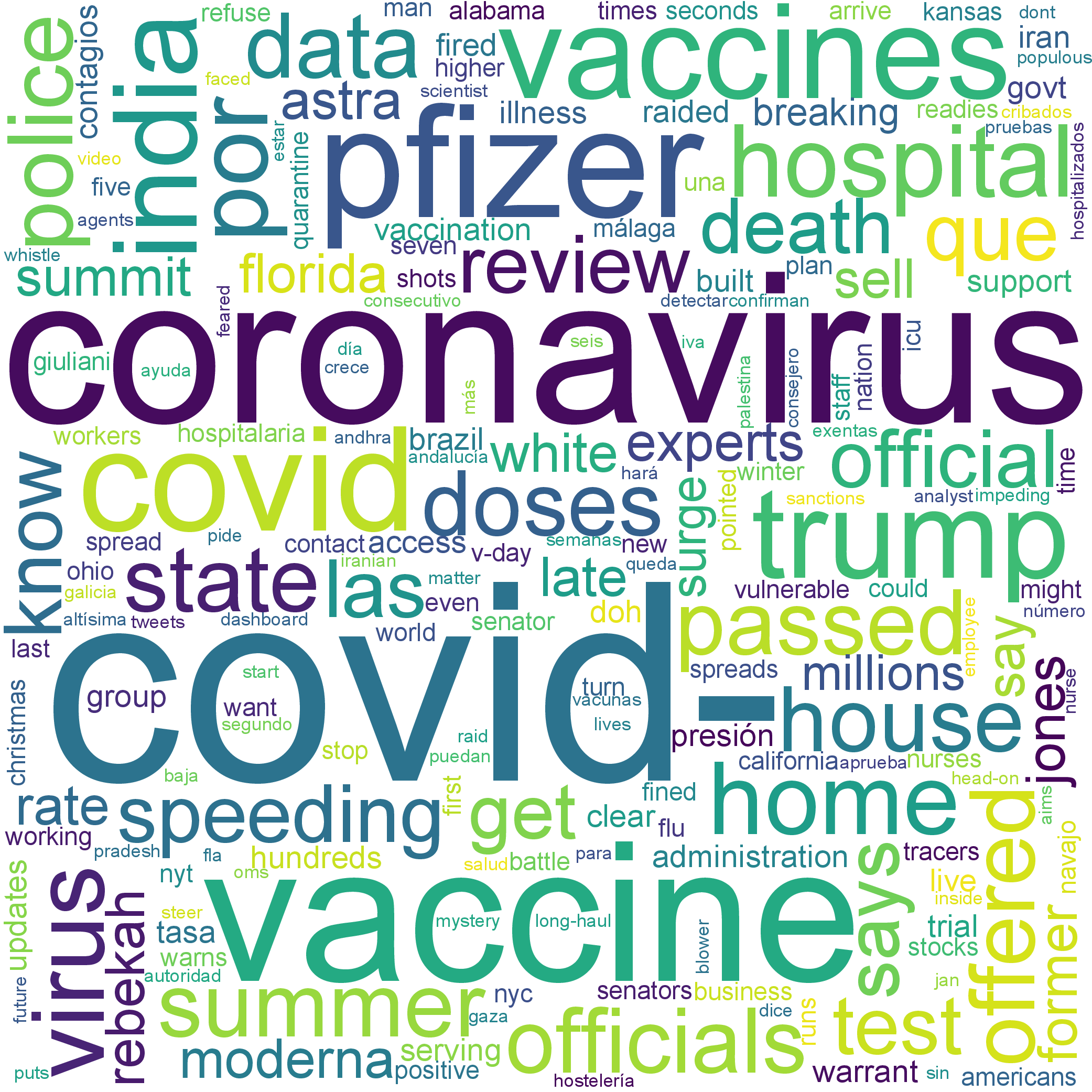}
    \end{tabular}
    \caption{Word clouds of 100 posts title from subreddit r/Coronavirus\protect\footnotemark at different times during COVID-19 pandemic (left to right: 01-02-2020, 01-05-2020 and 08-12-2020)}
    \label{fig:reddit-tudfe-coronavirus}
\end{figure}

To illustrate the capabilities of the developed application, let us present few examples, in two groups, in Figures~\ref{fig:reddit-tudfe-coronavirus} and \ref{fig:reddit-tudfe-music}. The wordclouds are build from posts related to a subreddit dedicated (or closest) to the searched topic. \textit{Reddit-TUDFE} allows to quickly check if, and how, a particular topic is covered. Note that similar examples can be derived for any other topic, while Reddit also shows potential in, for instance, building ontologies, or semantic graphs. However, this possibility is out of scope of this contribution.

In Figure~\ref{fig:reddit-tudfe-music}:
\begin{itemize}
    \item Left subfigure shows result for the phrase ``music'', a generic term, which is certainly discussed on Reddit. One may see particular genres: rock, pop, rap, relaxing, electronic, etc. 
    \item Middle subfigure displays results for phrase ``rock'', a bit narrowed, but still vague music-related (sub)topic, which is also present in Reddit, including artists/bands like: Rolling Stones, AC/DC, Led Zeppeling, Queen, Pink etc. 
    \item Right subfigure contains a strictly specific topic, i.e. the band ``The Beatles'', which is also widely covered on Reddit. Here one may see, among others, individual band members: John Lennon, Paul McCartney, Ringo Starr, and George Harrison.
\end{itemize}

Another example is summarized in Figure~\ref{fig:reddit-tudfe-coronavirus}.

In this figure, one can notice a clear shift of focus on the subreddit r/Coronavirus\footnotetext{https://www.reddit.com/r/Coronavirus/}
at different times of COVID-19 pandemic~\cite{abd2021comprehensive} (phrase ``coronavirus'' is skipped). 
Figure \ref{fig:reddit-tudfe-coronavirus} (left) (before 01-02-2020) shows that the main interest concerned the phrases a.o.: ``confirmed'' (the number of confirmed infections), ``Wuhan'' and ``Chinese'' (the geographical origins of the fist reported infections\cite{leung2020first}). 
Figure~\ref{fig:reddit-tudfe-coronavirus} (middle) (before 01-05-2020) displays that the main phrases changed to: ``deaths'' (due to COVID-19 infection) and ``lockdown'' (the preventive measures against the spread of the virus). Figure~\ref{fig:reddit-tudfe-coronavirus} (right) (before 08-12-2020, i.e. near the first vaccine invention) shows the general interests in phrases like: ``vaccine'' and ``pfizer'' (the company to invent the vaccine~\cite{badiani2020pfizer}).

Note that analysing the evolution of thematic ecosystem is just one of possible applications of the \textit{Reddit--TUDFE} tool. Most importantly, it quickly allows checking whether given topical domain contains live (evolving over time) information.
 
\section{Concluding remarks}

This work provides evidence that Reddit is a robust, but underutilized, resource for information retrieval and knowledge capture, in almost any field of interest. Based on performed exploratory analysis, the following answers to the research questions formulated at the beginning of this work can be stipulated:

\begin{itemize}
    \item \textbf{RQ1}: Reddit offers publicly available data, which can be easily retrieved with Pushshift API.
    
    \item \textbf{RQ2}: Most popular techniques for Reddit information processing are: text embeddings, neural networks, and graph networks.
    
    \item \textbf{RQ3}: Reddit is trending in scientific research as more and more articles using it are published every year.
    
    \item \textbf{RQ4}: Reddit covers the majority (79\%) of topics that appear in Global Google Trends, sustaining the claim that Reddit is a robust source of knowledge about ``everything trendy''.
    
    \item \textbf{RQ5}: Reddit is most commonly used in tandem with Twitter.
\end{itemize}

These conclusions render Reddit a perfect candidate for future research -- especially the presence of graph networks among common research methods and high coverage of popular trends. Finally, this analysis and the \textit{Reddit--TUDFE} tool provide solid foundation for future research on Reddit and its potential in information retrieval. 

\section*{Acknowledgement} 
This work has been supported in part by the joint research project “Novel methods for development of distributed systems” under the  agreement  on scientific  cooperation  between  the  Polish Academy of Sciences and Romanian Academy.

\bibliographystyle{siamplain}
\bibliography{references}
\end{document}